\documentclass[12pt,preprint]{aastex}

\newcommand{\ha}{H$\alpha$}

\newcommand{\oiiir}{[O~{\scriptsize III}]~$\lambda$5007}
\newcommand{\hii}{H~{\scriptsize II}}

\newcommand{\teff}{$T_{\mathrm{eff}}$}

\newcommand{\cho}{{\sc\footnotesize  CHORIZOS}}
\newcommand{\stb}{{\sc\footnotesize Starburst99}}
\newcommand{\rv}{$R_{5495}$}
\newcommand{\ecc}{$E(4405-5495)$}
\newcommand{\logoh}{$12 + \log(\mathrm{O/H})$}  
\newcommand{\hrd}{ Hertzsprung--Russell diagram}
\newcommand{\mbol}{$M_\mathrm{bol}$}
\slugcomment{Published in   the Astronomical Journal}
\shorttitle{The young stellar population of NGC 4214.I}
\shortauthors{\'Ubeda et al.}

\begin{document}

\title{The young stellar population of NGC 4214 as observed with 
 HST. I. Data and methods.\footnote{Based on observations made with the NASA/ESA 
{\it Hubble Space Telescope,} obtained at the Space Telescope Science
 Institute, which is operated by the Association of Universities for 
 Research in Astronomy, Inc., under NASA contract NAS 5-26555.}}

\author{ Leonardo \'{U}beda, Jes\'us  
Ma\'{i}z Apell\'aniz\footnote{Affiliated with the Space Telescope Division of the 
European Space Agency, ESTEC, Noordwijk, Netherlands.} , \& John W. MacKenty}
\email{lubeda@stsci.edu, jmaiz@stsci.edu, mackenty@stsci.edu}
\affil{Space Telescope Science Institute, 3700 San 
Martin Drive, Baltimore, MD 21218, U.S.A.}

\begin{abstract} 
We present the data and methods that we have used to perform a detailed UV--optical study of the
nearby dwarf starburst galaxy NGC~4214 using multifilter HST/WFPC2+STIS photometry.
We explain the process followed to obtain high--quality photometry and astrometry of the
stellar and cluster populations of this galaxy. We describe the procedure used to transform magnitudes
and colors into physical parameters using spectral energy distributions. 
The data show the existence of both young and old stellar populations that 
can be resolved at the distance of NGC 4214 (2.94 Mpc) 
and we perform a general description of those populations.

\end{abstract}


 \section{Introduction}

NGC~4214 is a nearby dwarf IAB(s)m galaxy \citep{Vauc91} in the low--redshift
CVn I Cloud \citep{Sand94} located at a distance of $2.94 \pm 0.18 $ Mpc \citep{Maizetal02a}.
The galaxy is moderately metal--deficient \citep{KobuSkil96} with \logoh\ between $8.15$ and $8.28$. 
The optical nebular morphology of NGC 4214 \citep{MacKetal00}  shows three differentiated components:
(1) two large \hii\ star--forming complexes, known in the literature as NGC 4214--I (or
NW complex) and NGC 4214--II (or SE complex); (2) a number of isolated 
fainter knots scattered throughout the field; (3)  an extended, 
structurally amorphous, Diffuse Interstellar Gas (DIG) 
surrounding the two main complexes and some of the isolated knots; see Figure~\ref{fig-1}.
The individual \ha\ knots have been identified by several authors in 
the past and  \citet{MacKetal00} have provided a more standardized 
nomenclature with 13 different units, which we adopt for this work.

What follows is a  summary of this galaxy's  morphology. Refer to
Figure~\ref{fig-2} for guidance. 
NGC 4214--I is the largest \hii\ complex in the galaxy and  the
one with the most intricate morphology.  It includes several star clusters and
has a complex \ha\ structure dominated by the presence of two cavities or
intensity minima.
I--As   is a massive young Super Star Cluster 
(SSC) located off--center in a heart--shaped 
\ha\ cavity. The \ha\ emission surrounding I--As consists of a number of knots joined by 
filaments.
I--Bs is a scaled OB association (or SOBA, 
\citealt{Hunt99}, \citealt{Maiz01b}): a massive cluster that, as opposed to an SSC, does not show 
a marked central concentration. I--Bs is also located in the middle 
 of an elongated \ha\ cavity, though not as clearly defined as the I--As one. 
NGC~4214--II is the second largest complex in the galaxy (in angular size) and
harbors the regions with the highest peak \ha\ intensity. Its morphology is 
quite different from that of NGC~4214--I. In the first place, there is no 
dominant SSC but several smaller clusters which are responsible for the ionization 
of the gas. Also, there are no large cavities present and the clusters are 
located very close to (or at the same position as) 
the most intensely \ha\ emitting knots
(\citealt{Maizetal98}). 
II--A and II--B are the two brightest \ha\ knots in NGC 4214--II. They are
both located on top of  clusters  which, at the resolution of WFPC2, appear to be
centered at the same positions.
NGC~4214--IIIs is a compact continuum source with weak associated \ha\ 
emission. It is either an older cluster or the nucleus of the galaxy, as suggested 
by the study of \citet{Faneetal97} in the I band and the FUV. 
NGC 4214--IVs has a similar appearance to NGC 4214--IIIs and it is  
probably another  intermediate-age massive cluster.
 \citet{MacKetal00} provides a more detailed description. 

The high intensity of the recent star and cluster formation rate in NGC~4214 combined with its 
proximity and its low foreground extinction make this galaxy an excellent target to test several
subjects of astrophysical interest  regarding young stellar populations. 
Nearby galaxies, and in particular NGC~4214,  
provide ideal laboratories  to test how stars form, how 
star formation is triggered, and details of how galaxies assembled. 
The study of nearby galaxies like NGC~4214 is rather important because 
most of the information that we infer about  high--redshift  galaxies relies on
what we observe in galaxies in the local universe. Therefore, understanding
nearby galaxies is of extreme importance to comprehend what
is going on in more distant ones.
NGC~4214 is a low--metallicity galaxy, and this gave us the possibility
to study the physical conditions   in an environment  of current astrophysical interest.

In this paper we present the data and the methods that we used to analyze it.
We have built detailed photometric lists with the most accurate astrometry. 
Of particular interest is the new  IDL code \citep{Maiz04} that we used
to  transform
observed magnitudes and photometric colors into actual 
stellar and cluster physical  parameters.
We also make a brief description of  its stellar  population.
 
In the accompanying 
paper [\citep{Ubedetalb06}, hereafter Paper~II]  we will study 
the ratio of blue to red (B/R)
supergiants, the initial stellar mass function (IMF), 
 the variable extinction across the galaxy, and the properties of
  its young-- and
intermediate--age cluster populations. This division is needed due to the vast amount 
of data and results that we have gathered, and to  the intricate  
structure of the object of study: NGC~4214 presents a number of clusters
in different stages of evolution, individually
resolved stars, and several regions of star formation, characterized by the presence
of complex nebulosities. H~{\scriptsize II} regions of 
different morphology  can be clearly seen  in our
images, showing the distinct nature of this interesting galaxy.

The present paper is organized as follows: In Section 2  we provide a description 
of our observations and data reduction, with emphasis on the creation 
of the photometric lists and the consistency tests. In Section 2.5 we 
introduce the method used to translate the observed magnitudes and colors into 
physical parameters. In Section 3 we provide a general description of the stellar populations
visible in our images. A brief summary is provided in  Sections 4.

\section{Observations and data reduction}

\subsection{WFPC2 Observations and reduction}

Wide Field and Planetary Camera 2 (WFPC2) is a two-dimensional 
imaging photometer  which 
consists of three CCD cameras (WF) with a spatial sampling of 
$0\farcs1$ per pixel and a smaller CCD camera (PC) with 
$0\farcs046$ per pixel.

The WFPC2  images were acquired   in two     programs: 
 6569  (P.I.: John MacKenty) and 6716 (P.I.: Theodore Stecher).  Figure~\ref{fig-1}  
shows the location of the WFPC2 fields for the   programs that we 
have used. All images include complexes  NGC~4214--I and NGC~4214--II. 
Table~\ref{tbl01} lists  all the NGC~4214 WFPC2 data that we have 
used for this paper.  The images available in each program can be 
summarized as follows:  6716, UV  and $UVI$; 6569,  narrow--band
(\oiiir, and \ha) + broad--band $UVI$. 
The WFPC2 observations of program 6569 were obtained on 
22 July 1997, and   those  of program 6716 were obtained on 29 June and 
9 December 1997.
For proposal 6569, the pointing of the telescope  was chosen in order
to minimize charge transfer efficiency (CTE) effects, since in this way 
no area of interest was separated from its collecting point by low signal 
areas \citep{MacKetal00}. In the case of  proposal   6716, the location 
was chosen in order to capture the most interesting regions of NGC~4214
with the PC, a camera that provides a higher resolution. These images 
allow us to obtain  accurate single star photometry  not easily achievable 
from the ground due to the very  complex diffuse emission and crowding.

The standard WFPC2 pipeline process takes care of the basic data 
reduction (bias, dark, flat--field corrections). The reduction of the processed 
frames  was performed using the PSF--fitting package {\sc\footnotesize  HSTphot} 
\citep{Dolp00a}, which yielded calibrated magnitudes in the VEGAMAG system corrected for
charge--transfer efficiency (CTE) for all filters.
We ran {\sc\footnotesize  HSTphot}  twice: first using a flag that estimates the 
sky locally, and later using the flag  that refits the sky during photometry. In both runs
we  used the default aperture corrections for each filter
provided by  {\sc\footnotesize  HSTphot}. 
The adopted magnitude for each star is the mean obtained from the two executions, and the errors 
in the magnitudes were estimated from the internal  errors in each run. 
The consistency of the photometry is analyzed in Section 2.3.

The WFPC2 CCD windows suffer from a  time--dependent
contamination  which primarily affects UV observations and is  
negligible at optical wavelengths. The contaminants are largely removed 
during periodic warmings (decontaminations) of the camera, and the effect upon
photometry is both linear and stable and can be removed using values regularly 
measured in the WFPC2 calibration program \citep{McMaWhit02}. We performed 
contamination correction for images obtained with filters F170W and F336W.
The presence  of saturated objects was analyzed in all the images, and
we found that cluster I--As appears saturated in  filter F555W of proposal 6716
and in  filters  F336W, F555W, and F702W  of  proposal 6569.
Cluster IIIs appears saturated in all  images obtained using filter F555W.
In  Section  2.4 we explain how we obtained the magnitudes for these clusters.

\subsubsection{Astrometry}

The geometric distortion of the WFPC2 detectors is well known \citep{CaseWigg01}, 
allowing for  precise relative astrometry. However, the absolute astrometry has 
two problems: First, the Guide Star Catalog, which is used as a reference on board 
{\it  Hubble Space Telescope}, has typical errors of $\sim 1\arcsec$ \citep{Russetal90}. 
Second, if only one point in one of the four WFPC2 chips can be established as a 
precise reference using an external catalog, the average error in the orientation 
induces an error of $\sim 0\farcs03$ in a typical position in the other three chips. 
In our case, two objects   which are present in the 
 WFPC2 fields  are included in the USNO--A2.0 catalog as entries 1200--06870167 
 and 1200--06870199. We used the VizieR service \citep{Ochsetal00} to obtain 
 their  coordinates  using the J2000 epoch and equinox, and we adopted
 object 1200--06870167 as the astrometric reference point.
This star is located
 in the center of region VIn which lies outside our studied field 
 but which  can be clearly seen in Figure~2 in   \cite{MacKetal00}.
 1200--06870199 is located in a very crowded region filled with many 
objects  in our high resolution images, and therefore, its  coordinates
correspond to a blend of point sources, making it useless for our purposes. 
The procedure that  we followed  to establish a
uniform coordinate system had three steps: 
First, we corrected for the difference in plate scale for each filter with respect to 
F555W using the parameters provided by \citet{Dolp00a} (note that 
the   tasks {\sc\footnotesize  mosaic} and {\sc\footnotesize  metric} 
within the {\sc\footnotesize   IRAF/STSDAS} package\footnote{The Image Reduction and Analysis Facility
({\sc\footnotesize   IRAF} ) is distributed by the National Optical Astronomy Observatories, which is operated
by the Association of Universities for Research in Astronomy, Inc., under
cooperative agreement with the National Science Foundation. 
{\sc\footnotesize   STSDAS} , the Space 
Telescope Science Data Analysis System, contains tasks complementary
to the existing {\sc\footnotesize   IRAF}  tasks. We used version 3.3.1 (2005 March) for our analysis.}, 
 use the non--wavelength--dependent
Holtzmann solution, which can introduce $\sim 0\farcs1$ errors for FUV data; 
errors are up to an order of magnitude smaller if only optical data are involved).
Second,  we corrected for  the geometrical distortion and built mosaics  of the four
 WFPC2 fields using the {\sc\footnotesize  mosaic} utility,   which is included in the standard 
 {\sc\footnotesize   STSDAS}  package for WFPC2 data analysis. We rotated all mosaiced images 
 in order to make them share a common orientation, keeping in mind that 
 this contributes to the astrometric accuracy due to subpixel rebinning. 
Third, using 1200--06870167  as a reference object, we corrected for the general 
displacement in right ascension and declination and compared the positions of the 
stars of several  regions of NGC~4214 using images obtained under different filters.
 As expected, coordinates differed by a few hundredths of an arcsecond, which we 
 take to be the precision of our absolute astrometry.

\subsubsection{Cross--correlation }
The field--of--view used for our analysis is shown in  Figure~\ref{fig-2} and
its dimensions are:  875 pc $\times$ 972 pc or  $ 61 \farcs 4 \times 68 \farcs 3$.
The region of the galaxy depicted by our field--of--view
 is clearly the most interesting one to study,
because it includes the most prominent star--forming regions, as well as
some clusters of importance.
 This field--of--view  represents the overlapping section of several  
images   obtained with HST/WFPC2 and HST/STIS (see next section). 

The last step in the data reduction, the registration of the images obtained 
under different programs, was somewhat complicated. We developed 
custom--made IDL scripts  which allowed us to cross--identify stars in  the 
different bandpasses by positional matching. In order to join the lists obtained
in different filters, we built two main photometric lists using F336W and F814W
as reference filters.  The lists are referred to here as {\sc\footnotesize  LIST336} and
{\sc\footnotesize  LIST814}.  Even though the individual lists used in the cross--correlation
were obtained with the same instrument, the orientation of the cameras 
was different for each list and the resolution and S/N used to image each
object was also different. To combine the final proposal--6569 list with the 
final proposal--6716 list with filter
F336W as reference, we used a cross--correlation procedure with the 
following characteristics:
In the most crowded regions in our images (NGC~4214--I) we only considered 
objects from proposal 6716 because  these images have a  higher resolution. 
Outside that region, the cross--correlation could yield several cases: (a) an 
object found in proposal 6569 but not found in proposal 6716, (b)  an object
found in both proposals, (c)  an object found in proposal 6569 that  corresponds
to two or more objects in proposal 6716. Our final list included all objects in 
case (a); if the object fell in case (b), we would consider the data from proposal
6569 due to its higher S/N, and get rid of its companion from proposal 6716. 
Finally,  if the object fell in case (c) we would consider those objects from 
proposal 6716 due to the higher resolution of the PC and get rid of its companion 
from proposal 6569.
With all the objects that were not correlated, we built another list
using filter F814W as reference.  {\sc\footnotesize  LIST336} has 14\,389 objects
and  {\sc\footnotesize  LIST814} has 21\,614 objects. 
%
Both lists of objects 
 ({\sc\footnotesize  LIST336} and  {\sc\footnotesize  LIST814}) are
 available in the electronic edition of this paper.
Tables~\ref{tbl02} and \ref{tbl03} contain the first five lines of those lists as a sample.  
The tables include the equatorial coordinates ($\alpha$, $\delta$) for each object
and their corresponding PSF (WFPC2 filters) or aperture (STIS) photometry.

We analyzed the photometric errors as a function of magnitude for 
all the considered filters. The photometric uncertainties are determined for 
each star based upon statistical errors, sky determination errors and aperture 
correction errors. We found that some points clearly exceed the  statistical errors, 
especially
in the F555W and F814W  bands: these are stars  deeply embedded 
in the  \hii ~regions where line emission significantly   enhances the local background.
Figure~\ref{fig-3}  presents six  plots of the magnitude errors as a function
of magnitudes F336W, F555W, and F814W obtained using our 
photometry lists from proposals 6716 and 6569.

\subsection{STIS Observations and reduction}

We summarize the STIS  data sets analyzed for this study in Table~\ref{tbl01}. 
All the UV observations employed the STIS near--ultraviolet (NUV) 
Cs$_2$Te Multi-Anode Microchannel Array (MAMA) detector. This detector has a 
field of view of $25" \times  25" $  and a pixel size of $\approx 0 \farcs 02468.$ 
The  images were obtained from proposal 9096 (P.I.: Jes\'us  
Ma\'{i}z-Apell\'aniz) and they show the two targets, regions 
NGC4214--I and NGC4214--II, with higher resolution than the WFPC2 images.

Each target was imaged using two different orientations
 shown in Figure~\ref{fig-4}. This   is an  RGB mosaic
which we made by assigning the  F336W image to the blue pixel values,
the F555W+\oiiir~ images to the green pixel values, and
the  F814W+\ha~ images to the red pixel values.
 Filter  F25CN182  (F25CN270) provides 
medium--band width imaging with a central  wavelength of 1820 \AA\ (2700 \AA) and
 an FWHM of  350 \AA\ (350 \AA).
The STIS data  were reduced by the STScI  calibration pipeline, via the 
on--the--fly reprocessing
which uses the best calibration reference files 
available at the time of retrieval from the HST archive. 
This used the latest geometric 
 distortion correction taken from   \citet{MaizUbed04}.

Comparing  Figures~\ref{fig-1} and~\ref{fig-4}     we see that 
the STIS field--of--view  is situated within the WFPC2
images of NGC~4214.  We obtained the  coordinates 
 of all objects in our 11 STIS images using a customized version
 of the   {\sc\footnotesize   DAOFIND} task in the
  {\sc\footnotesize   DAOPHOT} software package \cite{Stet87}
  running under   IDL   and then we performed
 aperture photometry on  all the images.  We cross--correlated the 11 STIS
 photometric lists with the list obtained from the WFPC2 images using 
 F336W as reference filter ({\sc\footnotesize  LIST336}).  We found that 
 a few  stars 
  in the WFPC2 list have two or more counterparts in the STIS list. 
This   is the result of the fact that STIS has a higher resolution than 
WFPC2. 
At the  distance  of NGC~4214 (2.94 Mpc), 
 the problem of multiple unresolved systems  
 is unavoidable. For example:   
 \citet{Maizetal05} study the problem derived 
from stellar multiplicity in the determination
of initial mass functions. They present results for
Trumpler 14, a  massive young cluster  in the Carina Nebula
that contains at least three 
very--early O--type stars.
If this cluster were located at a distance similar to that of NGC~4214, it
would appear as a point source even though
we know for certain that it is a multiple system. 
We use this result to justify our
approach to solve this issue, by including
 a combination of objects as a single object in our final list.
%
%

\subsection{Completeness tests}
The detailed analysis  of  NGC~4214 data (such as 
the study of the B/R ratio and the derivation of an IMF)  
require  a quantitative evaluation of completeness
of the photometric data.  
We therefore conducted several experiments in 
which we added artificial 
stars with known magnitude and position
to the images and then attempted to recover them using the
 same finding procedures 
that we used for the real stars. 
This was done using routines within the {\sc\footnotesize  HSTphot} suite of codes. 
The artificial stars were given 
random magnitudes and colors in the range $14 \le $F336W $  \le    26$ 
 and $-2.0 \le$ F336W -- F555W $   \le    4.0$. 
Approximately 50\,000 fake stars were added in each image in all four chips. 
{\sc\footnotesize  HSTphot}  generates a grid of artificial stars
which are distributed according to the flux of the image, so that
crowded areas would contain more fake stars than less crowded ones. 
Then the artificial stars were detected and measured using the 
same algorithms that we 
used for  the observed data. 
To be recovered, a star  must have survived the fitting process, having been 
found automatically and it must have passed the cross--identification test.
The ratios of recovered to inserted stars
in the different magnitude bins give directly  the completeness factor
as a function of magnitude.
In Figure~\ref{fig-5}, we 
plotted the results of the artificial star tests for filters F336W, F555W,
 and F814W in proposals 6569 and 6717  and
 we observe that, at the bright end, the completeness is about 95\% 
or  higher in all images.  For fainter 
magnitudes, the completeness falls off more quickly with increasing 
magnitude in the images obtained in 
 proposal 6716 than in those obtained in proposal 6569, as expected 
from the shorter exposure times of the former.

The next step was to map the \hrd ~as a  finely spaced grid, and to calculate 
 the completeness values at  each point. In order to do this, we
used  \cho\ (see Section 2.5)  to obtain two useful relationships:  
 the relation between
F336W -- F555W and \teff;
and the relation between F336W -- F555W and \mbol\ for 
$E(4405-5495) = 0.1 $ mag  which is a typical value of the extinction 
in NGC 4214 as  estimated by \citet{Leitetal96}.
 This 
procedure was only used to estimate the completeness values 
which we list on Table~\ref{tbl04}, not 
to determine stellar or cluster physical properties.
A separate  independent study of completeness for the STIS 
data was not necessary, because we matched
all the stars in the STIS lists to stars in   {\sc\footnotesize  LIST336}. Therefore,
the completeness tests conducted  for  {\sc\footnotesize  LIST336}
 include both WFPC2 and STIS data. Since the STIS images go deeper
 than the WFPC2 data, our results are limited by the depth of the WFPC2 images.

In order to check for the consistency of our data, we used the fact that
NGC 4214 was visited twice for proposal 6716: first on  June 29, 1997    and later
on December 09,  1997. The orientation of the cameras in the sky was 
different for both visits as can be clearly seen in Figure~\ref{fig-1}. It is 
important to note that the idea behind this proposal was to obtain the best
resolution of regions NGC~4214--I and NGC~4214--II and that is  
why  those complexes are centered in the PC.
The fields of view that correspond to both visits in  proposal 6716  contain
a significant amount of overlap, allowing us to test the internal consistency 
of the photometry in order to check for possible systematic uncertainties introduced
by the CTE or contamination corrections.
For each of the four filters in this proposal we obtained two
photometric lists, one for each visit to NGC~4214.  We then performed a 
statistical test on all the objects  which were identified as the same  star in
the two lists.   We calculated the distribution of $(m_1 -  m_2) /   
\sqrt{  \sigma_{1}^{2} + \sigma_2^{2}} $, where $m_i$ and $ \sigma_i$   are
the magnitude and its uncertainty in list $i$.  This distribution 
closely resembles a normal distribution with zero mean and a standard deviation of unity 
for all the filters used, as we show 
in the histograms in Figure~\ref{fig-6}. The total number of points (N) considered 
for the histograms, the average value of  $  \Delta m /  
\sqrt{  \sigma_{1}^{2} + \sigma_2^{2}}    $ and its standard 
deviation are given in Table~\ref{tbl05}.
The same study with filters F25CN182 and  F25CN270  was performed
for our STIS lists. Again, we obtained histograms  that closely resemble
a normal distribution with zero mean and standard deviation of unity, 
showing the agreement between  different photometric lists.  

The completeness values obtained in this section 
 will allow us to better estimate the actual number
of blue and red supergiants in the galaxy. We will also use them
 to infer
the initial mass function as we explain in detail in Paper~II.

\subsection{Cluster photometry}

The {\sc\footnotesize  HSTphot} package  \citep{Dolp00a} 
was developed specifically to obtain stellar photometry of point 
sources on  images obtained  with HST WFPC2. Our WFPC2 and STIS images contain some
 objects  (I--As, I--Es, IIIs, and  IVs) 
which are known to be unresolved stellar clusters, and which appear  as extended
sources rather than point--like objects.  The position of these clusters is 
shown in Figure~\ref{fig-2},  and their corresponding  blowups are in Figure~\ref{fig-7}.
Other clusters are resolved into stars and we also
analyzed them using aperture photometry.
These clusters are organized in two groups: large complexes
(which include smaller clusters): I--A, I--B, and  II;
and resolved clusters: I--Ds, II--A, II--B, II--C, II--D, and II--E.
Their positions in the galaxy are shown 
in  Figure~\ref{fig-2}.

We performed detailed aperture photometry of these clusters,  
carefully selecting the aperture radii and the sky apertures.
We used the closest  possible radii values to those provided by  \citep{MacKetal00}.
In all cases we included  more than 90\%  of the light from each
cluster, with the exception of cluster I--As, which presents an extended halo. 
See Paper~II for a full discussion.
We used custom made IDL
codes to perform this task on WFPC2 and STIS CCD images. The values of the aperture 
correction (in magnitudes) that we used to calculate  the aperture correction for the STIS images 
are taken from  \citet{Prof03} which refer to point sources.
When the cluster was bright enough and no confusion with nearby sources was apparent, 
we measured the $JHK_s$ photometry from 2MASS data using the same
apertures that we had used with  the WFPC2 and STIS images. 
Clusters IIIs and IVs have their  $JHK_s$  magnitudes listed in the 2MASS catalogue.
We compared the profile   fitting and  the aperture photometric measurement provided by 2MASS with our own measurements to determine 
an aperture correction. This correction was required  to compare
the 2MASS  magnitudes with the WFPC2 ones, given the
different spatial resolutions.
The complete photometry 
of all the   clusters is summarized in Tables~\ref{tbl06} and \ref{tbl07}.

\subsubsection{Unresolved clusters}

Images from proposal 6716 show cluster I--As  on the PC of the WFPC2, while those of proposal 
6569 picture this region  on the WF3. 
This cluster is the brightest feature in NGC~4214 at optical wavelengths, 
and its pixels appear saturated in some images
(filter F555W in proposal 6716; filters  F336W, F555W, and F702W in proposal 6569.)
This object can also be seen in some of our STIS images. Unfortunately,
images obtained during programme 6716 on December 1997 show a bad column passing through 
cluster I--As; in order to get rid of this problem, we interpolated the number of counts in each pixel along
the column using the pixels from  two adjacent columns of the chip. 
Aperture photometry was obtained using an aperture radius of 9 PC pixels which is
equivalent to $5.84 $ pc at a distance of 2.94 Mpc, or $0\farcs 41. $ 
The adopted equatorial coordinates of the cluster are given in  Table~\ref{tbl08}.
Once we obtained the
magnitudes in all the available images, we calculated a weighted mean for each filter, 
using their own errors as weights. 

A small cluster can be seen between I--A and I--B: cluster I--Es.
It  was imaged  in proposal 6716  images using  filters F170W, F336W, F555W,  
and F814W on the PC.  It was also captured  in all the images from proposal 6569 on the WF3
camera.  This object can also be seen in some of our STIS images.
Some images of this cluster show a bad column and we solved  this 
problem using  the same technique as  we did for cluster I--As.

Cluster IIIs  can be  found in proposal 6716  images  in filters F170W, F336W 
and F814W on the WF4. It appears as a saturated object in filter F555W. It was also found in 
all the images from proposal 6569 on the WF3 camera, but the central pixels are  saturated in  image F702W.
On images from proposal 6716, the cluster appears  very near  the  vignetted field between the
PC and the WF4. For this reason, we thought that the calculated magnitudes might be affected
and we discarded all of them except for the magnitude obtained with filter
 F170W, and we relied on the magnitudes obtained with images from proposal 6569.
Aperture photometry was obtained using an aperture radius of 12 WF pixels which is
equivalent to $17.1 $ pc at a distance of 2.94 Mpc. Once we obtained the
magnitudes in all the available images, we calculated a weighted mean for each filter. 
This  object  is present in the 2MASS All--Sky Catalog of Point Sources  \citep{Cutretal06} and is
located in a relatively isolated region, 
which provided us with useful magnitudes in the $J$, $H$ and $K_s$ bands. Nevertheless, 
an aperture correction was required in order to compare those magnitudes with the WFPC2 ones, given the
different spatial resolutions. The field--of--view of the available NICMOS images
does not cover the clusters analyzed in this paper. 

Cluster IVs falls on the vignetted field between the
WF3 and the WF4 on the proposal 6716 images, making it impossible to calculate
its magnitude in those images.
This object is clearly visible in   proposal 6569 images in filters F336W, F555W, F702W and
F814W pictured on the WF3.
Aperture photometry was obtained using an aperture radius of   11.40  pc or $0\farcs 8. $ 
Cluster IVs is also listed in the  2MASS  Catalog. We retrieved its  $J$, $H$ and $K_s$ magnitudes and
corrected them for aperture effects as we did for cluster IIIs.
The complete photometry 
of these  clusters is summarized in Table~\ref{tbl06}.

\subsubsection{Large complexes of smaller clusters or SOBAs}

We performed aperture photometry  for complexes   I--A, I--B, and II
using    aperture radii very similar to the ones that that \citet{MacKetal00} 
use to describe the clusters. 
We centered cluster I--A at a position slightly displaced from cluster I--As.
Whenever we found bad columns in our images, we corrected them
using the same procedure that we employed for cluster I--As. 
Some of our images have some  saturated pixels within an  aperture radius  of 150 PC pixels;
those filters were discarded. We were able to obtain the photometry of this cluster in  
images with filters F170W, F336W, F814W, F25CN182, and F25CN270.

Another cluster which is part of the large complex NGC~4214--I is 
cluster I--B, which is  located to the south--east of cluster I--A. We  
used an aperture radius of 42.18 pc  ($ 2\farcs 96 $). 
We were not able 
to use the images from proposal 6569 because this cluster lies very close to  the
vignetted section of the chips. Some of the images in proposal
6716 have bad columns across the cluster. 

The other large nebular structure in the galaxy is NGC~4214--II. 
This complex  part of NGC~4214 is very well imaged on the WF4 chip  from proposal 6716,
on the WF2  chip from proposal 6569, and on some of our  STIS images. 
We used  
an  aperture radius  of 200 PC (or 91 WF) pixels to perform the photometry. We encountered  some 
bad columns extending across this cluster, specially through clusters II--A, and  II--C.

Using the images provided by the  2MASS All--Sky Survey, we 
performed aperture photometry in filters $J$, $H$, and $K_s$ for clusters I--A, I--B, and II.
We used the centers and aperture radii given  in Table~\ref{tbl08}.
To calibrate our measurements, we used the magnitudes given by the 
2MASS All--Sky Catalog for clusters IIIs and IVs, which were calculated using
a $ 4\farcs 0 $ aperture radius.
Since the sky in these images is very noisy (the number of counts in the sky 
is of the order of counts in the clusters), we obtained big errors in our 2MASS photometry.
The complete photometry 
of these  clusters is summarized in Table~\ref{tbl06}.

\subsubsection{Resolved clusters}

Cluster I--Ds is composed of a small group of stars located towards the north of cluster
I--A, and has only a small amount  of gas in its surroundings \citep{MacKetal00}. 
A small aperture of 20 PC pixels (12.97 pc)  was used  for the photometry.
This cluster falls out of the field--of--view of our STIS images.
It is located near the vignetted area in the WF3 camera chip from proposal
6569. We also found it on the PC in one of the visits from proposal 6716 and
on the WF4 in the other visit. 

We performed a detailed analysis of the structure of
cluster II by  examining its five  components 
(II--A, II--B, II--C, II--D, and II--E.)
These fairly well--resolved clusters are embedded within several \ha  ~regions
as can be observed comparing Figures~\ref{fig-2} and ~\ref{fig-4}.
To perform the aperture photometry, we 
followed the same routine as for cluster II, taking into account
the bad columns in all cases. The aperture radii for these clusters are
50, 37, 57, 32, and 35 PC pixels respectively or 
32.49, 23.94, 36.91, 20.81, 22.66 pc respectively.
The complete photometry 
of these  clusters is summarized in Table~\ref{tbl07}.

\subsection{SED fits using {\bf\sc\footnotesize  CHORIZOS}}

Our images provide photometric magnitudes  in various filters,
which  represent the main observable. To translate these observables into
fundamental stellar/cluster  parameters -- e.g. effective temperature  (\teff), 
bolometric magnitude (\mbol), and cluster age --
we used  \cho ~\citep{Maiz04}, an IDL package
that fits an arbitrary family of spectral energy  distribution 
models (SEDs) to multi--color photometric data. 
The fundamental parameters are determined  simultaneously 
by a likelihood--maximization technique from the photometric colors
calculated from the measured magnitudes.
The associated errors for these parameters are 
computed from the resulting likelihood multidimensional array.  
We applied \cho\ to stars and clusters separately,
using different sets of theoretical models.

\subsubsection{Application to stars}
We applied \cho\ to our observed stellar magnitudes to measure the effective temperature, 
monochromatic color excess [ \ecc, see e.g. \citealt{Maiz04} for its relationship to $E(B-V)$ ],
and (when possible) extinction law of the stars   in our sample. 
For the stars we
used as input SEDs low--gravity Kurucz models with $\log(Z/Z_{\sun})= -0.5$.
The stars in {\sc\footnotesize  LIST336} were treated in  the following way:

\begin{description}
\item[six and five filters ]  All the stars with six and five 
 observed magnitudes were processed through \cho\ first. For these objects \teff, \ecc, and the extinction
law were left as free parameters. The extinction law was chosen among the \rv--dependent family
of \citet{Cardetal89}, and the MC laws of \citet{Missetal99} and \citet{GordClay98}.
  

\item[four filters ]  For the objects with only four observed magnitudes (three independent
colors), the extinction law was restricted to three values in three different runs: First we ran 
the code using  
an SMC--like  extinction law; as a second experiment, we used a  Galactic--extinction
law with the widely used value    $R_{5495}= 3.1$, and for the last experiment  
we fixed the  value $R_{5495}= 5.0$.
We compared the values of  the goodness of the fit $(\chi^2) $ for each star
 and kept the results  that correspond to the
lowest value of  $\chi^2. $ 
For this set of objects we had three independent colors and, 
having fixed the   extinction law, only two parameters to be determined:
\teff  ~and \ecc.
  
\item[three filters ]  The remaining objects with only three observed 
magnitudes were processed through \cho\ leaving only \teff\ and \ecc\ as free 
parameters and using the extinction law of \citet{Cardetal89}, with  
$R_{5495}= 3.1$. For this set of objects, we had only two independent colors and we had to 
fix at least one parameter. This left us with zero degrees of freedom but this is the best 
that can be done with only three observed magnitudes. 

\end{description}

It is well known from studies of our own and
other galaxies that
the extinction law can vary in small  (less than 1 pc)
spatial scales (see   \cite{Ariaetal05} for a recent study). We decided to fix $R_{5495}= 3.1$
to the objects with three and four filters, because this value is very close to
the  mean values of $R_{5495}$ calculated for stars with five and six filters, where
we left $R_{5495}$  as a free parameter in the fit, and because $R_{5495}= 3.1$
is the standard value measured in the Galaxy.

Objects in {\sc\footnotesize   LIST814} have magnitudes in three WFPC2
 filters:  F555W, F702W, and F814W,
with which we built two photometric colors: F555W--F702W and F555W--F814W.
In this case, the maximum allowed number of free parameters is two but, for this
specific filter combination, there are strong color degeneracies that hamper the simultaneous
determination of \teff\ and \ecc. Therefore, we decide to use our  extinction results  (see Paper~II) by 
restricting \ecc\ to the range  $-0.05 \leqslant  E(4405-5495) \leqslant 0.25  $ and by fixing
$R_{5495}= 3.1$, as explained above.

In order to assess the goodness of the fit we rely on the value of $\chi^2$
provided by \cho ~for each star.  All objects with    $\chi^2 \leq  4.0 $  were 
considered as good fits throughout our work.  
We performed a detailed analysis of those objects from lists with six and five 
filters with $\chi^2 >  4.0$. An inspection of the output plots which shows  the
theoretical spectrum overplotted on the measured magnitudes, reveals in 
some of these cases that one or two of the observed  magnitudes are not in
agreement with the rest. This may be due to several reasons, such as strong nebular
contamination in F555W or unresolved multiple systems. Those cases were analyzed on a 
one--to--one basis. In all these cases, we accordingly removed one or two magnitudes
from the list (e.g. F555W for a star with strong nebular contamination or F814W for an early--type star
with an apparent late--type companion) and reran \cho\ with the new set of magnitudes. 


The output of \cho\  that corresponds to
the stellar components of NGC~4214 will be used 
to analyze various aspects of this galaxy: In this paper we present
two Hertzsprung--Russell diagrams from which we infer the galaxy's stellar population.
In Paper~II we will implement these results
to study the variable extinction throughout    our field--of--view,
to calculate   the blue to red supergiant ratio, and to
determine the initial mass function slope using 
the  most direct and reliable method which  is based on counts of stars as a function
of their luminosity/mass.

\subsubsection{Application to clusters}

The extended sources in our sample (both resolved and unresolved) were 
treated in a different way.
\cho\ includes precalculated \stb\ \citep{Leitetal99} cluster models, where 
the intrinsic parameters are the cluster age and its metallicity and the external
parameters are the same as in the stellar case: \ecc\ and $R_{5495}$.
These models are tabulated for a \citet{Salp55}  stellar initial mass function
of 1 to 100 $M_\sun$.
The exact low mass limit of the  IMF  is not important for our age determinations
because the integrated colors of clusters  with ages younger than $10^8$ 
years are dominated by 
stars with masses greater than 1 $M_\sun$, but it is relevant
for the determination of the total mass.

For all our clusters
we used \stb\ models  of integrated stellar populations
with metallicity $Z/Z_{\sun}= 0.4$, appropiate for NGC~4214
 \citep{KobuSkil96}, and
ages of the models varying  between  $\log({\rm age/yr}) = 6.0$ 
and $\log({\rm age/yr}) = 9$.
The   available  reddening values \ecc ~are in the range $0.0 - 5.0.$
For the extinction law, we considered  
the $R_{5495}-$dependent family of 
 \citet{Cardetal89} (Galactic extinction); the average LMC and 
LMC2 laws of  \citet{Missetal99}   (extinction of the LMC), and the SMC
law of  \citet{GordClay98}  (extinction of the SMC). 

The comprehensive study of the \cho\ output  is presented in Paper~II, where
we address several issues such as the determination of clusters age, mass,
and massive stellar constituents. We analyze the different solutions for each cluster
and we present arguments to choose the best possible model that fits
our data in each case. In  Paper~II, we will  also discuss the problems that arise when 
using large and small aperture radii.

\section{General description of the observed stellar populations}

With the output values of  \cho ~from the stellar analysis, we can build a 
theoretical \hrd  ~for NGC~4214.
These diagrams allow us to compare
our observations  with stellar evolution models and to  
interpret them  in physical terms such as mass, age, composition etc. 
We  used sets of  evolutionary tracks provided by \cite{LeSc01}. 
These use the entire set of non--rotating Geneva stellar evolution models 
covering masses from $0.8$ to $120 M_{\sun}$.

NGC~4214 is moderately metal deficient, with an abundance of 
 $12 + \log(\mathrm{O/H}) = 8.15 - 8.28$ \citep{KobuSkil96} or about $Z = 0.34 Z_\odot  =  0.006 $
using the results provided by  \cite{Massey03} for the MCs and the Galaxy.
We bracketed the data between  theoretical models with  
metallicities  $Z=0.004$  and $Z=0.008$, 
which are the closest available  to the estimated  metallicity of NGC~4214 $(Z\approx 0.006).$ 

These theoretical models provide the luminosity, age,  and effective temperature 
of  representative evolutionary points. We calculated seven 
isochrones  by interpolating the bolometric magnitudes as a function
of the logarithm of the age along each mass track in steps of  $1-5$
million years (ZAMS, 1Myr, 2Myr, 3Myr, 5Myr, 10Myr, and 15Myr.)
The Hertzsprung--Russell diagrams of NGC~4214 are presented in 
Figures~\ref{fig-8} and \ref{fig-9}. 

We have separated the stars in two diagrams:   
Figure~\ref{fig-8} is a plot of objects from {\sc\footnotesize  LIST814} and Figure~\ref{fig-9} 
is a plot of objects taken from  {\sc\footnotesize  LIST336}. Instead of representing individual 
stars, we  display contour plots to show the varying density of stars 
in the [ $\log(T_{\mathrm{eff}}) $, $M_{\mathrm{bol}}$ ]   plane also 
referred to as Hess diagrams.  The contour
 levels correspond to stellar counts of 10, 15, 20, 40, 50, 100, 300, 600, 1200,
and 2000 objects per HRD element  in Figure~\ref{fig-8},  and 5, 10, 15, 20, 40, 50, 75, 100, 
150, and 200 objects per HRD element in Figure~\ref{fig-9}.  Each HRD element 
measures 0.0875 mag in 
the $M_{\mathrm{bol}}$  axis and 0.0075 in the $\log(T_{\mathrm{eff}}) $
axis.  Individual stars are shown 
as small crosses in those regions of the diagrams where the density was 
lower than 10 and 5 stars respectively .

\begin{figure}
\includegraphics[width=0.99\textwidth]{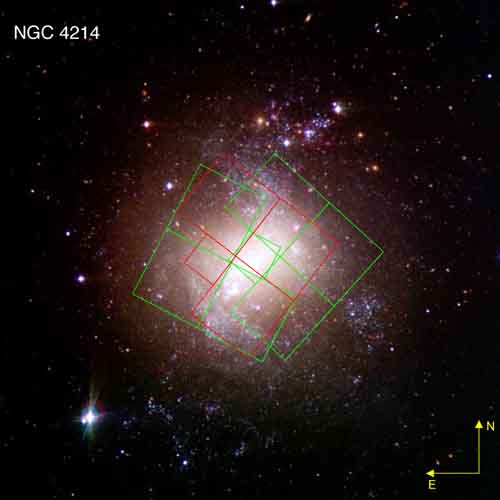}
\caption{$9\arcmin \times 9\arcmin$ (7.7 kpc $\times$ 7.7 kpc) 
image of NGC 4214 from an RGB combination 
of public {\it B, V} and {\it R} band images obtained with the Isaac 
Newton Telescope.   HST/WFPC2 fields (green)  are overplotted for 
proposal 6716. The left  field corresponds to the second visit (09 Dec 1997) 
and the right  field corresponds to the first  visit (29 Jun 1997). 
The HST/WFPC2 field from proposal 6569 is displayed in red (center field).  }
\label{fig-1}
\end{figure}

\begin{figure}
\includegraphics[width=0.9\textwidth]{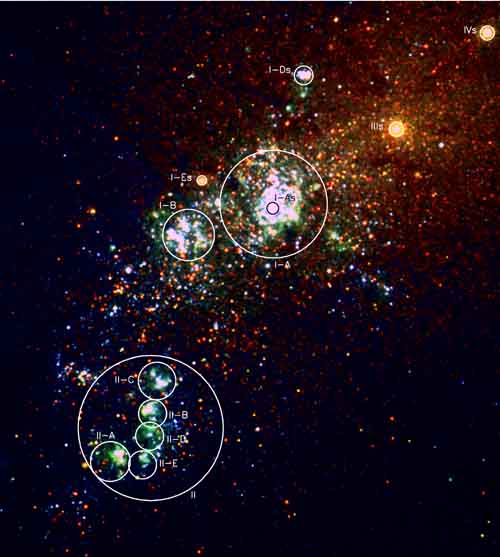}
\caption{Mosaic of NGC 4214 obtained using images in filter F336W (blue),
F555W ~(green), and F814W ~(red). 
We show and label the cluster circular apertures.
The  orientation  is north pointing up and east pointing to the left. 
The field dimensions are  875 pc $\times$ 972 pc or  $ 61 \farcs 4 \times 68 \farcs 3$.
}
 \label{fig-2}
\end{figure}

\begin{figure}
\includegraphics[width=0.9\textwidth]{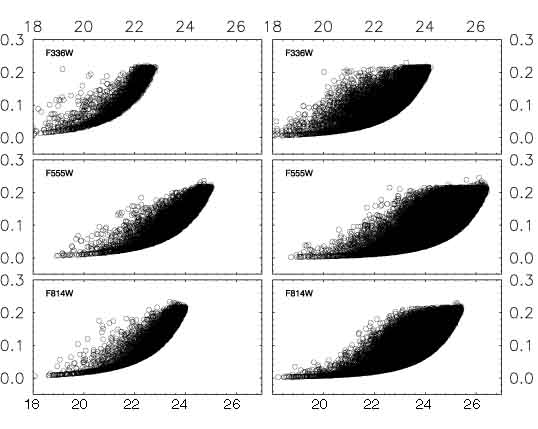}
\caption{ Residuals in magnitudes for the WFPC2
filters F336W, F555W, and F814W. The results in the left column 
correspond to the lists from proposal  6716.
The plots in the right column are those from proposal 6569.}
\label{fig-3}
\end{figure}

\begin{figure}
\includegraphics[width=0.9\textwidth]{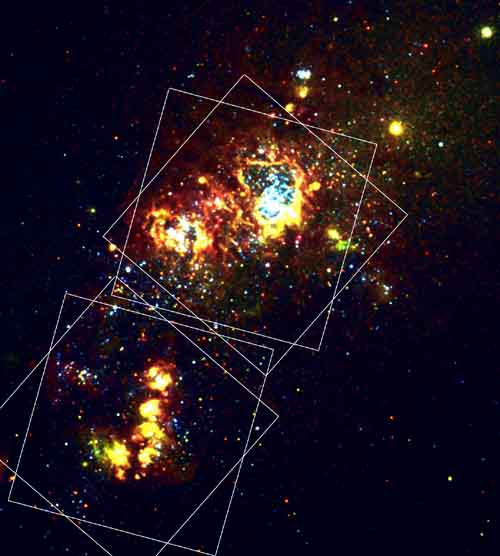}
\caption{HST/STIS fields drawn  over a WFPC2 color mosaic of NGC 4214.
Each field corresponds to one visit.
The mosaic was built  using original images in filters F336W (blue), 
F555W+\oiiir~(green), and  F814W+\ha~(red).
The field shown is the same as in Figure~\ref{fig-2}.
 }
 \label{fig-4}
\end{figure}

\begin{figure}
\includegraphics[width=0.9\textwidth]{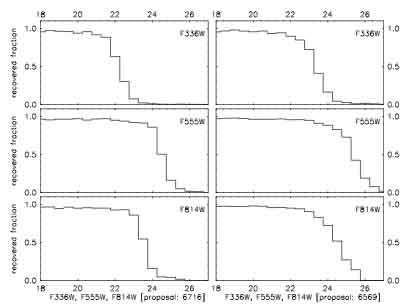}
\caption{ Completeness levels of the WFPC2 photometry based on artificial 
star tests performed using   {\sc\footnotesize  HSTphot} 
\citep{Dolp00a}. The  fraction of recovered artificial stars is plotted
against the magnitude for filters F336W, F555W, and F814W.
 The results in the left column 
correspond to the tests performed on images from proposal  6716.
The plots in the right column are the tests performed on images
 from proposal 6569. }
\label{fig-5}
\end{figure}

\begin{figure}

\includegraphics[width=0.7\textwidth]{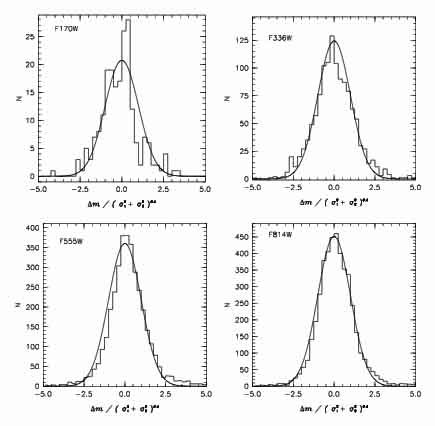}
\caption{Histograms for the measured magnitude difference normalized 
by the experimental uncertainty ( $  \Delta m /   \sqrt{ \sigma_{1}^{2} + \sigma_{2}^{2} } $ ). 
Here we are comparing magnitudes of stars  observed in two different 
visits during proposal 6716. See text for details about the method.   
The black  curve shows the expected normal distribution.
\label{fig-6}}
\end{figure}

\begin{figure}
\includegraphics[width=0.9\textwidth]{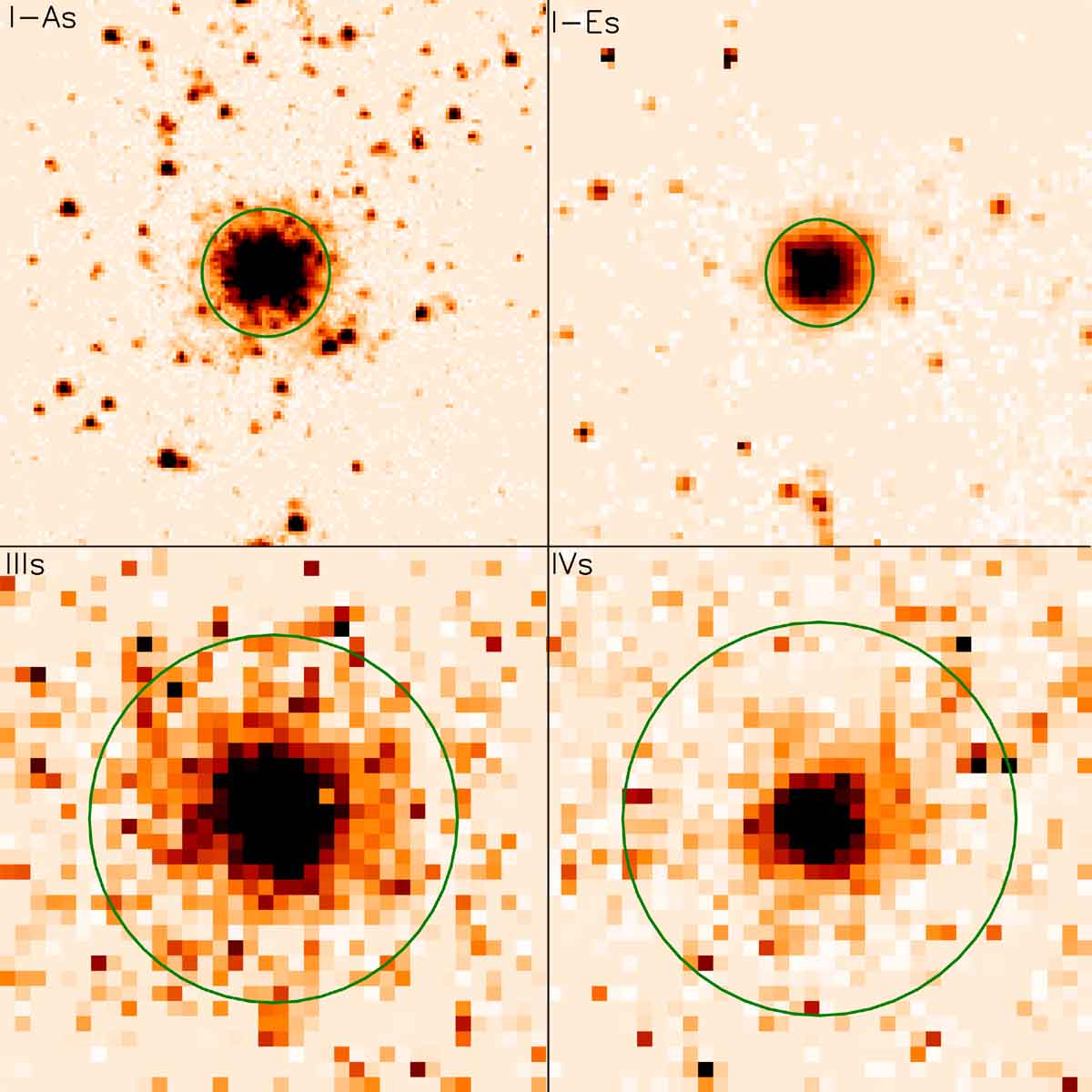}
\caption{HST images of star clusters  I--As, I--Es, IIIs, and  IVs.
The images have been scaled in order to use the same linear size, with the field
being $50$ pc  $\times 50$ pc in all cases. North is up and east is left. The circles depict the
apertures used for our photometry.
Cluster  I--As is shown on a STIS image in filter F25CN270. We used a WFPC2/PC image in filter F555W to 
show cluster I--Es. Clusters IIIs and IVs are depicted on WFPC2/WF F814W images.
These images clearly show the different pixel size (and resolution) of the three cameras that we have 
used.     }
 \label{fig-7}
\end{figure}

\begin{figure}
\includegraphics[width=0.9\textwidth]{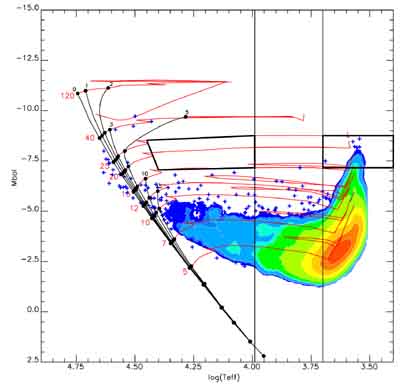}
\caption{ Hess diagram for all stars in {\sc\footnotesize  LIST814} (the main photometric
list using F814W as reference filter) with calculated \mbol and 
$T_{\mathrm{eff}}$  obtained from the \cho\ output.  The contour levels 
correspond to stellar counts of 10, 15, 20, 40, 50, 100, 300, 600, 1200, 
and 2000 objects per element.  Each element measures 0.0875 mag in 
the $M_{\mathrm{bol}}$  axis and 0.0075 in the $\log(T_{\mathrm{eff}}) $
axis. Individual stars are shown  as small crosses in those
regions of the diagrams where the density was lower than 10 stars.  The
red lines represent the evolutionary tracks for stars in the mass--range 5, 
7, 10, 12, 15, 20, 25, 40, and 120 $ M_{\sun}$. The solid black lines
represent the isochrones calculated for ages 0, 1, 2, 3, 5, 10, and 15 Ma. 
This diagram also shows two polygons
between tracks labeled 15 and 25  $ M_{\sun}$. These are the regions 
that we will use to count blue and 
red supergiants in Paper~II.  }
\label{fig-8}
\end{figure}

\begin{figure}
\includegraphics[width=0.9\textwidth]{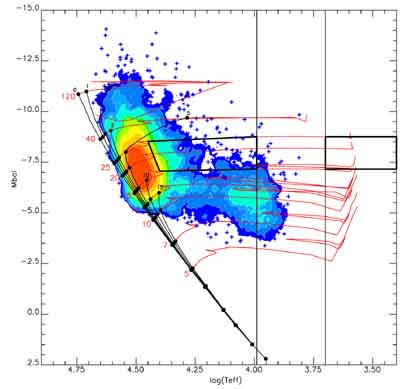}
\caption{ Hess diagram for all stars in {\sc\footnotesize  LIST336} 
 (the main photometric
list using F336W as reference filter) 
 with calculated \mbol  and  $T_{\mathrm{eff}}$   
obatined from the \cho  ~output.
The contour levels correspond to stellar counts of  5, 10,15,20,40,50,75,100,150, and 200 objects
per element.  Each element measures 0.0875 mag in 
the $M_{\mathrm{bol}}$  axis and 0.0075 in the $\log(T_{\mathrm{eff}}) $
axis.
 Individual stars are shown  as small crosses in those regions of the diagrams 
 where the density was lower than  5 stars.  The red lines represent the 
 evolutionary tracks for stars in the mass--range 5, 7, 10, 12, 15, 20, 25, 40, 
 and 120 $ M_{\sun}$. The solid black lines represent the isochrones
 calculated for ages 0, 1, 2, 3, 5, 10, and 15 Ma.
This diagram also shows two polygons
between tracks labeled 15 and 25  $ M_{\sun}$. These are the regions 
that we will  use to count blue and 
red supergiants  in Paper~II.}
\label{fig-9}
\end{figure}

As a dwarf irregular galaxy, NGC~4214 is an interesting
laboratory where the process of star formation can be studied. This is
a young galaxy which is still processing gas into stars. It is a relatively gas--rich system
with a young stellar population of bright, blue stars and \hii ~regions. The star formation
regions are concentrated along a bar structure surrounded by a large disk of gas. A second, less
intense, region of recent star formation, is located off the northern tip of the bar at a distance of
about 2 kpc from the galactic center but is not covered by the HST data (see Figure~\ref{fig-1}).
Because the ISM is readily ionized by the intense UV radiation from 
young stars in \hii\ regions, \ha\ is often used as an  indicator of the 
presence of recent star formation.  
Figure~\ref{fig-4} pictures
the regions of star formation through the \ha\ tracer. 

Figures~\ref{fig-8} and \ref{fig-9}  allow us to analyze the stellar population of NGC~4214.
Figure~\ref{fig-8}  shows the objects from  {\sc\footnotesize  LIST814}. The most striking feature
 present in this plot is the crowded clump in the red part of the
 diagram. This feature shows a great concentration of stars in the range
 $ 3.50 \leqslant   \log(T_{\mathrm{eff}}) \leqslant 3.75$.  It is composed 
 of the red giant  locus, the asymptotic giant branch, 
 and, along its blue extent, of intermediate
 age  blue loop stars.
 Most probably,
these stars represent a mixed--age population that has evolved 
off the zero--age main sequence.  
A group of red supergiants  stars can be seen at the tip of the red plume. 
These are stars with initial masses $\gtrsim  40 M_{\odot}$ which 
had evolved past the main sequence and into the RSG phase.
  One noticeable fact
is that some of the stars are located to the right of the extreme 
of the evolutionary tracks. This fact has been observed by  \cite{MassOlse03} while 
studying  RSGs in the Magellanic Clouds. 
The evolutionary tracks do not go far enough to the right
(cool temperatures) to produce the RSGs that are actually observed. 
The same problem was mentioned by \cite{Massey03} for Galactic RSGs.
\cite{Leveetal05} present a new effective temperature scale for Galactic 
RSGs  by fitting MARCS stellar atmosphere models \citep{Gustetal75, Plezetal92} which include an 
improved treatment of molecular opacity to 74 Galactic $(Z=0.020)  $ 
RSGs of known distance. 
Their main result is that RSGs appear to be warmer than previously thought.
This effect shifts the stars to the left in the diagrams (by $\sim 300-400$ K)
making them coincide with the end of the tracks.

Figure~\ref{fig-9}   is characterized by a blue plume of stars. 
This is the realm of the H--burning (main--sequence) massive stars.
The locus of these stars lies above the  $5 M_{\odot}$ evolutionary track and extends
up to $M_{\mathrm{bol}}  \approx -14. $ It is well populated in the range 
$\approx 12 M_{\odot} - 60 M_{\odot}$, indicating some ongoing star formation.
In the figure we can clearly see a group of objects lying along the main sequence,
indicating that these are young objects. However, for large masses, the mean locus of
the position of the stars is located to the right (lower temperatures) of this
sequence. This is likely to be caused by a combination of two effects: the difficulty
of observing ZAMS stars at high mass (because they are in highly extinguished regions)
and the existence of unresolved stellar systems (which also has the consequence of 
producing a number of points above the $120 M_{\odot}$ track). 
The other feature in this diagram is the group of stars located  between the  
$5 M_{\odot}$  and the $20 M_{\odot}$ evolutionary tracks, and with 
$\log(T_{\mathrm{eff}}) \lesssim 4.25$. These objects are evolved massive stars
with ages greater than 15 Myr.

We agree with  \cite{Droz02} in that  
stars appear to populate the \hrd\  of NGC~4214 throughout all 
of the major phases of stellar evolution,  leading  one  to conclude  
that the star formation was more or less continuous in recent times,
although there could have  been some recent starbursts present. 
Most of the brightest stars in the galaxy are concentrated towards the two 
main regions of star formation, which are NGC~4214--I and NGC~4214--II.
These are regions characterized by different morphology, which  
\cite{MacKetal00}  interpret as an evolutionary or aging trend. The stellar components
of NGC~4214-II are co-spatial with the gas, and individual massive  stars
can be seen in different stages of stellar evolution. As we show in Paper~II, this
is the youngest (age $\approx$ 2 Myr) region in the galaxy.
On the other hand, NGC~4214--I  (age $\approx 3-5$  Myr)  
shows some signs of cluster evolution, with
evacuated cavities produced by winds and supernova explosions. Younger
stars tend to concentrate around the main two clusters (I--A and I--B)
in this part of the galaxy.
Concentrations of bight stars are also detected outside the central area, 
outlining a bar structure  which is a characteristic of NGC~4214. 
The crowded clump in the red part of the
 diagram in Figure~\ref{fig-8} suggests that the galaxy possesses a 
 significant  underlying
 intermediate--age /  old  population.
 This fact is reinforced by the prominent red halo observed in Figure~\ref{fig-1}.
The analysis of \cite{Maizetal98}  shows that the older population
contributes to around 50 \% of the optical continuum. This population
does not contribute to the UV continuum, but dominates the emission
in the IR. All this means that the present burst is not the first episode of star formation
in the nucleus of NGC~4214.

\section{Summary and conclusions}
In this paper we present the HST/WFPC2 and HST/STIS observations
of the well--resolved nearby galaxy
NGC~4214. We obtained   PSF photometry of the stellar
components and   aperture photometry of some interesting clusters using high
resolution images. We explain how we managed to translate observed
quantities such as magnitudes and photometric colors  into
physical parameters such as \teff, and \mbol ~for stars, and age for clusters, using
 \cho ~\citep{Maiz04}.

The stellar content of NGC~4214 has been discussed using 
two  [ $\log(T_{\mathrm{eff}}) $, $M_{\mathrm{bol}}$ ] 
Hertzsprung--Russell diagrams: one shows the stars from a  list
with filter F336W as reference, and the other
 was built with filter F814W as reference.
The first diagram shows most objects  on the main--sequence, while
the second diagram  is characterized by a crowded concentration of
stars  in the range
 $ 3.50 \leqslant   \log(T_{\mathrm{eff}}) \leqslant 3.75$.

In Paper~II  we will present the following  results:  
the blue to red supergiant ratio, an estimate of the initial mass
function, a study of the variable extinction throughout the galaxy,
and a detailed analysis  of the cluster population.

\acknowledgments
We want to thank  Rupali Chandar, Claus Leitherer, and  Henry Ferguson 
for reading the first draft of the paper and making 
very useful  comments.
We would like to thank the 
  referee  (Dr. Richard de Grijs) for the many suggestions that have helped 
improve the manuscript. 
Support for this work was provided by NASA through grants GO--06569.01--A, 
GO--09096.01--A, GO--09419.01--A, and 
AR--09553.02--A from the Space Telescope Science Institute, Inc., under NASA contract 
NAS5--26555.
This research has made use of the VizieR catalogue access tool, CDS, Strasbourg, France.
This publication makes use of data products from the Two Micron All Sky Survey, which
 is a joint project of the University of Massachusetts and the Infrared Processing and 
 Analysis Center/California Institute of Technology, funded by the National Aeronautics 
 and Space Administration and the National Science Foundation.

Facilities: \facility{HST(WFPC2)}, \facility{HST(STIS)}.


\clearpage

\clearpage

\clearpage

\clearpage

\clearpage

\clearpage

\clearpage

\clearpage

\clearpage

\clearpage
\begin{deluxetable}{cllll} 
\tabletypesize{\scriptsize}
\tablecaption{Journal of observations, with filters and exposure times
 for WFPC2 and STIS images. The dates of the observations are given in the main text.    \label{tbl01}}
\tablewidth{0pt}
\tablehead{ \colhead{Proposal} & \colhead{Filter} &\colhead{Band} &\colhead{Dataset\tablenotemark{a}} &\colhead{Exposure Time (s) }   }
\startdata

	  & F336W \dotfill & WFPC2  {\it U}       & u3n80101m + 2m + 3m \dotfill & 260 + 900 + 900  \\
& F555W \dotfill & WFPC2 {\it V}        & u3n80104m + 5m + 6m \dotfill & 100 + 600 + 600  \\
\raisebox{2ex}[0pt]{6569}     	     & F702W \dotfill & WFPC2 wide {\it R}   & u3n80107m + 8m \dotfill  & 500 + 500  \\
{\rule [-2mm]{0mm}{0mm} }    & F814W \dotfill & WFPC2 {\it I}        & u3n8010am + bm + cm \dotfill & 100 + 600 + 600  \\ \tableline

{\rule [0mm]{0mm}{4mm} }   & F170W \dotfill & {\it UV}             & u4190101r + 102r + 201m + 202m & 400 + 400 + 400 + 400 \\
& F336W \dotfill & WFPC2  {\it U}       & u4190103r + 104r + 203m + 204m & 260 + 260 + 260 + 260 \\
 \raisebox{2ex}[0pt]{6716}  & F555W \dotfill & WFPC2 {\it V}        & u4190105r + 205m \dotfill    & 200 + 200  \\
{\rule [-2mm]{0mm}{0mm} }  & F814W \dotfill & WFPC2  {\it I}       & u4190106r + 206m \dotfill  & 200 + 200  \\ \tableline

{\rule [0mm]{0mm}{4mm} }  	  & F25CN182 \dotfill &       & o6bz02isq + 02iwq + 01afq \dotfill & 288 + 288 + 288  \\
& F25CN182 \dotfill&       & o6bz04r8q + 04raq + 03afq \dotfill & 288 + 288 + 288  \\
\raisebox{2ex}[0pt]{9096}	  & F25CN270 \dotfill &       & o6bz02j7q + 02jbq + 03awq\dotfill & 288 + 288 + 288  \\
{\rule [-2mm]{0mm}{0mm} } 	  & F25CN270 \dotfill  &       & o6bz04ruq + 04ryq \dotfill              & 288 + 288  \\

\enddata
\tablenotetext{a}{STScI identifiers.}
\end{deluxetable}


\begin{deluxetable}{lcccccccc} 
\tabletypesize{\scriptsize}
\tablecaption{  Photometry of stars in {\sc\footnotesize  LIST336}. The photometric 
uncertainties are determined for 
each star based upon statistical errors, sky determination errors and aperture 
correction errors.    The complete lists is in the electronic version of this paper.   \label{tbl02}}
\tablewidth{0pt}
\tablehead{  \colhead{$\alpha$ (J2000)} &  \colhead{$\delta$ (J2000)} & \colhead{F336W} &  \colhead{F170W} &  \colhead{F555W} &\colhead{F702W}&\colhead{F814W} &\colhead{CN182}&\colhead{CN270} }

\startdata

                        12:15:37.802                                       &                            36:19:44.200                                       &                       $17.55 \pm  0.01$                                       &                                \dotfill                                       &                       $19.06 \pm  0.01$                                       &                       $18.90 \pm  0.01$                                       &                       $18.94 \pm  0.04$                                       &                                \dotfill                                       &                                \dotfill                                      \\
                            12:15:39.584                                       &                            36:19:35.376                                       &                       $17.66 \pm  0.04$                                       &                       $17.67 \pm  0.10$                                       &                       $19.08 \pm  0.07$                                       &                                \dotfill                                       &                       $19.03 \pm  0.04$                                       &                       $17.43 \pm  0.12$                                       &                       $17.22 \pm  0.11$                                      \\
                            12:15:40.561                                       &                            36:19:33.604                                       &                       $17.86 \pm  0.02$                                       &                       $18.00 \pm  0.10$                                       &                       $19.24 \pm  0.01$                                       &                                \dotfill                                       &                       $19.16 \pm  0.02$                                       &                       $18.04 \pm  0.14$                                       &                       $17.73 \pm  0.12$                                      \\
                            12:15:39.586                                       &                            36:19:29.245                                       &                       $17.87 \pm  0.01$                                       &                       $17.25 \pm  0.06$                                       &                       $19.57 \pm  0.05$                                       &                       $19.57 \pm  0.05$                                       &                                \dotfill                                       &                       $17.04 \pm  0.11$                                       &                       $17.18 \pm  0.11$                                      \\
                            12:15:40.574                                       &                            36:19:33.422                                       &                       $17.88 \pm  0.06$                                       &                       $16.78 \pm  0.08$                                       &                       $19.62 \pm  0.07$                                       &                                \dotfill                                       &                       $19.74 \pm  0.06$                                       &                       $17.28 \pm  0.12$                                       &                       $17.65 \pm  0.11$                                      \\

\enddata
\end{deluxetable}


\begin{deluxetable}{lcccc} 
\tabletypesize{\scriptsize}
\tablecaption{  Photometry of stars in {\sc\footnotesize  LIST814}. The photometric 
uncertainties are determined for 
each star based upon statistical errors, sky determination errors and aperture 
correction errors. The complete lists is in the electronic version of this paper.     \label{tbl03}}
\tablewidth{0pt}
\tablehead{  \colhead{$\alpha$ (J2000)} &  \colhead{$\delta$ (J2000)} & \colhead{F814W} &   \colhead{F555W} &\colhead{F702W} }

\startdata

        12:15:41.449                                       &                            36:18:59.382                                       &                       $18.80 \pm  0.00$                                       &                       $20.79 \pm  0.01$                                       &                       $19.42 \pm  0.00$                                      \\
                            12:15:39.384                                       &                            36:18:59.959                                       &                       $18.83 \pm  0.00$                                       &                       $21.16 \pm  0.01$                                       &                       $19.59 \pm  0.01$                                      \\
                            12:15:33.382                                       &                            36:19:25.804                                       &                       $18.94 \pm  0.00$                                       &                       $25.32 \pm  0.16$                                       &                       $24.81 \pm  0.22$                                      \\
                            12:15:32.202                                       &                            36:19:36.543                                       &                       $18.95 \pm  0.01$                                       &                       $20.90 \pm  0.04$                                       &                       $19.51 \pm  0.08$                                      \\
                            12:15:38.633                                       &                            36:20:40.951                                       &                       $19.06 \pm  0.00$                                       &                       $20.51 \pm  0.00$                                       &                       $19.18 \pm  0.00$                                      \\

\enddata
\end{deluxetable}

\begin{table}[htbp]
\caption{Completeness values for 
each track  from   \cite{LeSc01} and their corresponding main sequence lifetime ($\tau_{MS}$).
 \label{tbl04}  \vspace{5mm}}
\begin{tabular}{ l|cccccccccc  } \tableline\tableline
Track    ($M_{\odot}$) {\rule [-2mm]{0mm}{7mm} }
  &  7    &  10 & 12 & 15 & 20 & 25 & 40 & 60 & 85 & 120     \\ \tableline
  Completeness     
   &  0.041   &  0.263 & 0.497 & 0.705 & 0.855 & 0.899 & 0.929 & 0.946 & 0.953 & 1.000  {\rule [0mm]{0mm}{5mm} }  \\  
   $\tau_{MS}$    (Myr) 
   &  46.50  &  23.98 & 17.84 & 12.95 & 9.12 & 7.17 & 4.82 & 3.71 & 3.48 & 3.01 {\rule [-3mm]{0mm}{0mm} }    \\  \tableline
\end{tabular}
\end{table}


\begin{deluxetable}{lcccc} 
\tabletypesize{\scriptsize}
\tablecaption{Analysis of the consistency of the photometry
using two WFPC2 photometry lists from proposal 6716 (P.I.: Theodore Stecher).  \label{tbl05}}
\tablewidth{0pt}
\tablehead{  \colhead{ } & \colhead{F170W} &  \colhead{F336W} &  \colhead{F555W} &\colhead{F814W}  }
\startdata
N       & 189  & 1114  & 3219  & 4116   \\  
average\tablenotemark{a} & -0.120&-0.013  & 0.119 & 0.049  \\
$\sigma\tablenotemark{b}$ & 0.839 & 0.915 & 0.841 & 0.847  \\ 
\enddata

\tablenotetext{a}{Average value of $(m_1 -  m_2) /   \sqrt{  \sigma_{1}^{2} + \sigma_2^{2}} $.}
\tablenotetext{b}{Standard deviation of the  quantity in (a).}

\end{deluxetable}

\clearpage

\begin{deluxetable}{lccccccc} 
\tabletypesize{\scriptsize}
\tablecaption{Photometry of unresolved clusters and of large complexes \label{tbl06}}
\tablewidth{0pt}
\tablehead{  \colhead{Filter} & \colhead{Cluster I--As} &  \colhead{Cluster I--Es} &  \colhead{Cluster IIIs} &  \colhead{Cluster IVs} 
& \colhead{Cluster I--A} &  \colhead{Cluster I--B} &  \colhead{Cluster II}   }
\startdata
F170W   \dotfill  & 13.47 $\pm$ 0.07  & 18.82 $\pm$ 0.52  & 16.77 $\pm$ 0.15  & \dotfill    &11.81 $\pm$ 0.02  & 13.54 $\pm$ 0.03  & 12.45 $\pm$ 0.04\\
F25CN182  &  13.50 $\pm$ 0.02  &  19.27 $\pm$ 0.10  & \dotfill  & \dotfill  &11.99 $\pm$ 0.02  &13.60 $\pm$ 0.02&  12.91 $\pm$ 0.02   \\
F25CN270  &  14.00 $\pm$ 0.03   & 19.56 $\pm$ 0.11  &  \dotfill  & \dotfill  & 12.51 $\pm$ 0.01  & 14.17 $\pm$ 0.02 & 13.20 $\pm$ 0.02 \\
F336W  \dotfill &  14.46 $\pm$ 0.04  &  19.30 $\pm$ 0.03  & 16.82 $\pm$ 0.05  & 17.17 $\pm$ 0.03  & 12.96 $\pm$ 0.01  & 14.62 $\pm$ 0.01 &13.59 $\pm$ 0.01  \\
F555W  \dotfill & \dotfill  & 18.98 $\pm$ 0.05  & 16.59 $\pm$ 0.03  &17.30 $\pm$ 0.01  & \dotfill  &15.83 $\pm$ 0.23   & 14.37 $\pm$ 0.04   \\
F702W  \dotfill & \dotfill  & 18.27 $\pm$ 0.04&  \dotfill  & 16.97 $\pm$ 0.01 & \dotfill  & \dotfill    &   14.04 $\pm$ 0.08 \\
F814W  \dotfill & 15.98 $\pm$ 0.02  & 17.90 $\pm$ 0.03  & 15.67 $\pm$ 0.03  & 16.75 $\pm$ 0.02 & 13.93 $\pm$ 0.01 & 15.55 $\pm$ 0.01&  14.24 $\pm$ 0.01   \\
$J$ \dotfill & \dotfill  & \dotfill  &  14.74 $\pm$ 0.10 &15.93 $\pm$ 0.13   & 14.43 $\pm$ 0.50  &  15.69 $\pm$ 0.25 &  14.27 $\pm$ 0.63  \\
$H$  \dotfill  & \dotfill  & \dotfill  & 14.30 $\pm$ 0.11 &  15.55 $\pm$ 0.19 & 14.31 $\pm$  0.88 &15.05 $\pm$ 0.27  &13.65 $\pm$ 0.70   \\
$K_s$  \dotfill   &  \dotfill  & \dotfill  &13.87 $\pm$ 0.10   & 15.39 $\pm$ 0.20  &  13.54 $\pm$  0.66  & 15.27 $\pm$ 0.42  &    13.71 $\pm$ 1.02     \\
\enddata
\end{deluxetable}

\begin{deluxetable}{lcccccc} 
\tabletypesize{\scriptsize}
\tablecaption{Photometry of resolved clusters \label{tbl07}}
\tablewidth{0pt}
\tablehead{  \colhead{Filter} & \colhead{Cluster I--Ds} &  \colhead{Cluster II--A} &  \colhead{Cluster II--B} &  \colhead{Cluster II--C} 
& \colhead{Cluster II--D} &  \colhead{Cluster II--E}    }
\startdata
F170W   \dotfill  & 15.51 $\pm$ 0.03  &15.54 $\pm$ 0.18  & 15.26  $\pm$ 0.05  & 14.23 $\pm$ 0.04 &16.39 $\pm$  0.16&15.97 $\pm$ 0.13 \\
 F25CN182  &  \dotfill  &15.63 $\pm$ 0.06  & 15.50 $\pm$ 0.02  & 14.67 $\pm$  0.02 &16.66 $\pm$ 0.06 &16.12 $\pm$  0.04 \\
 F25CN270  &  \dotfill   &15.66  $\pm$ 0.06  &15.70 $\pm$  0.02    &15.07 $\pm$ 0.02&16.96 $\pm$  0.07&  16.33 $\pm$  0.04\\  
 F336W  \dotfill &  16.39 $\pm$ 0.01  &15.87  $\pm$ 0.01  &16.04 $\pm$ 0.01  & 15.52 $\pm$ 0.01& 17.30 $\pm$  0.02 &16.67 $\pm$  0.01 \\ 
 F555W  \dotfill &17.62 $\pm$ 0.02&16.46 $\pm$ 0.08  & 16.87 $\pm$ 0.03  & 16.76 $\pm$ 0.06 &18.12 $\pm$  0.12&17.75 $\pm$  0.06\\
 F702W  \dotfill & 17.54 $\pm$ 0.04  & 16.09  $\pm$ 0.07& 16.56 $\pm$  0.06  & 16.50 $\pm$  0.12&18.00 $\pm$  0.22& 17.48 $\pm$ 0.05\\
 F814W  \dotfill & 17.43 $\pm$ 0.03  & 16.46  $\pm$ 0.08  & 17.02 $\pm$ 0.01  & 16.72 $\pm$ 0.01&18.21 $\pm$  0.01&17.93 $\pm$  0.01\\
 \enddata
\end{deluxetable}

\clearpage
 
\begin{deluxetable}{lccccc} 
\tabletypesize{\scriptsize}
\tablecaption{The clusters astrometry and aperture radii.  \label{tbl08}}
\tablewidth{0pt}
\tablehead{  \colhead{\raisebox{-0.9ex}[0pt]{Cluster}  } & \colhead{ $\alpha$ (J2000)} & \colhead{ $\delta$ (J2000)} & 
\colhead{Radius} & \colhead{Radius}  \\
\colhead{} & \colhead{ $12^h 15^m +$ } & \colhead{ $36^{\circ} +$ } &  \colhead{["]}  &  \colhead{[pc]}  }
\startdata

      I--As  \dotfill  &39.54&      19  36.49& 0.41 &     5.84       \\ 
      I--Es  \dotfill &40.26&       19  39.94& 0.36 &   5.13         \\ 
      IIIs\dotfill   &38.27&        19  46.29& 1.20& 17.10           \\ 
       IVs \dotfill  &37.34&       19  58.20& 0.80 &  11.40          \\  
       
   I--A\dotfill   &  39.53      &  19   37.00  &    6.83    &97.33            \\
       I--B \dotfill  &  40.40  &  19   33.18  & 2.96 & 42.18           \\
        II \dotfill &   40.79    &  19  09.26  & 9.10&  129.68          \\ 
        
       I--Ds\dotfill   &39.22  &   19  52.96  & 0.91&  12.97          \\ 
      II--A \dotfill  &41.20    &   19   05.12& 2.28&  32.49          \\ 
      II--B \dotfill   &40.77&      19  11.06& 1.68&  23.94          \\ 
      II--C \dotfill   &40.72&      19  15.11& 2.59 &  36.91          \\ 
      II--D \dotfill   &40.80&      19   08.25& 1.46& 20.81           \\ 
      II--E \dotfill  &40.87&       19   04.71& 1.59 &  22.66          \\ 
\enddata
\end{deluxetable}

\end{document}